\documentstyle[12pt]{article}

\catcode`\@=11

\def\@normalsize{\@setsize\normalsize{15pt}\xiipt\@xiipt
\abovedisplayskip 14pt plus3pt minus3pt%
\belowdisplayskip \abovedisplayskip
\abovedisplayshortskip  \z@ plus3pt%
\belowdisplayshortskip  7pt plus3.5pt minus0pt}

\def\small{\@setsize\small{13.6pt}\xipt\@xipt
\abovedisplayskip 13pt plus3pt minus3pt%
\belowdisplayskip \abovedisplayskip
\abovedisplayshortskip  \z@ plus3pt%
\belowdisplayshortskip  7pt plus3.5pt minus0pt
\def\@listi{\parsep 4.5pt plus 2pt minus 1pt
            \itemsep \parsep
            \topsep 9pt plus 3pt minus 3pt}}

\def\underline#1{\relax\ifmmode\@@underline#1\else
        $\@@underline{\hbox{#1}}$\relax\fi}
\@twosidetrue
\relax

\catcode`@=12

\catcode`\@=11

\def\ps@headings{\def\@oddfoot{}\def\@evenfoot{}
\def\@oddhead{\hbox{}\hfill
        \makebox[.5\textwidth]{\raggedright\ignorespaces --\thepage{}--
\hfill {}}}}
\def\@evenhead{\@oddhead}
\ps@headings

\catcode`\@=12

\relax

\def\figcap{\section*{Figure Captions\markboth
        {FIGURECAPTIONS}{FIGURECAPTIONS}}\list
        {Fig. \arabic{enumi}:\hfill}{\settowidth\labelwidth{Fig. 999:}
        \leftmargin\labelwidth
        \advance\leftmargin\labelsep\usecounter{enumi}}}
 \relax
\def\tablecap{\section*{Table Captions\markboth
        {TABLECAPTIONS}{TABLECAPTIONS}}\list
        {Table \arabic{enumi}:\hfill}{\settowidth\labelwidth{Table 999:}
        \leftmargin\labelwidth
        \advance\leftmargin\labelsep\usecounter{enumi}}}
 \relax
\def\reflist{\section*{References\markboth
        {REFLIST}{REFLIST}}\list
        {[\arabic{enumi}]\hfill}{\settowidth\labelwidth{[999]}
        \leftmargin\labelwidth
        \advance\leftmargin\labelsep\usecounter{enumi}}}
 \relax

\catcode`\@=11

\def\@evenhead{\@oddhead}
\ps@headings

\relax

\newskip\humongous \humongous=0pt plus 1000pt minus 1000pt

\newif\ifdtup

\def\beq{\begin{equation}}
\def\eeq{\end{equation}}

\def\beqn{\begin{eqnarray}}
\def\eeqn{\end{eqnarray}}
\relax

\def\G2{{\; \rm GeV/}c^2}
\def\G{\; \rm GeV}

\def\dotx{\dotx{\dot\overline{x}}}

\relax

\begin{document}

\begin{titlepage}
\nopagebreak
\begin{flushright}

        {\normalsize     ITP--SB--92--33\\
                         June,~1992\\}

\end{flushright}

\vfill
\begin{center}
{\large \bf Integrable Superhierarchy  \\
                     of  \\
            Discretized
 2d  Supergravity}\footnote{Work supported in part by NSF Grant Phy 91-08054}

\vfill
        {\bf H.~Itoyama}\footnote{Address after July 1st: Physics Department,
 Faculty of Science, Osaka University, Toyonaka, Osaka, 560, Japan} \\

        Institute for Theoretical Physics,\\
        State University of New York at Stony Brook,\\
        Stony Brook, NY 11794-3840, USA\\

%\maketitle\thispagestyle{empty}

\end{center}
\vfill

\begin{abstract}
 The  hierarchical nonlinear super-differential
 equations  are identified which
 describe  universal behavior of the discretized model of $2d$ supergravity
recently proposed. This is done by first taking a double scaling limit
 of the super Virasoro constraints ( at finite $N$) of the model
  and by rederiving it from the  $\tilde{G}_{-1/2}$ constraint and the
 two reduction of
the super KP hierarchy discussed.
 The double-scaled constraints are  found to be described
 by a twisted scalar and a Ramond fermion.
\end{abstract}
\vfill

\end{titlepage}
  Matrix models have had  successes recent years
 in providing  us an opportunity to analyze nonperturbative effects
in string theory.
  A deep connection to  integrable system
has been revealed, which we hope  plays  pivotal roles in the subsequent
developments to come.

Somewhat surprisingly, there has been no substantial progress in
 constructing discretized models
 with supersymmetry which describe discretized $2d$-supergravity and/or
the one coupled to super-conformal matter.  Going through this tantalizing
period,  a model has recently been found \cite{AIMZ}
 whose  properties at the $k$-th critical point  agree
with the results obtained from the continuum formulation of the
$2d$-supergravity coupled to the $(2,4k)$ superconformal matter \cite{PZDHK}.
The partition function contains the integrations over
 Grassmann coordinates
$\theta_{i}$ as well as the ones over the
 eigenvalue
coordinates $\lambda_{i}$:
\beq
\label{eq:model}
  {\bf Z}_{N}[g_{p},\xi_{p}] =  \int {\displaystyle \prod_{i=1}^{N}
 d\lambda_{i} d \theta_{i}} {\bf \Delta} \left( \lambda_{i}, \theta_{i}
 \right)
e^{ - \frac{N}{\Lambda} {\displaystyle \sum_{i=1}^{N} {\bf V}
\left( \lambda_{i}, \theta_{i}; g_{p}, \xi_{p+1/2} \right)} }
\;\;\;.
\eeq
 Here, $ {\bf \Delta} \left( \lambda_{i}, \theta_{i}
 \right)$ is a supersymmetric generalization of the Vandermonde
determinant
 $ {\bf \Delta} \left( \lambda_{i},
 \theta_{i} \right) = {\displaystyle \prod_{i >j} }
 \left( \lambda_{i} -\lambda_{j}
- \theta_{i} \theta_{j} \right) $ and
$  {\bf V} \left( \lambda,
 \theta ; g_{m}, \xi_{m+1/2} \right) \equiv {\displaystyle
 \sum_{p=0}^{\infty}
g_{p} \lambda^{p} + \sum_{p=0}^{\infty} \xi_{p+1/2} \theta \lambda^{p} }$.
The degree of the potential at the lowest critical point $(k=1)$
alone tells us that the model contains more than just gravitational dressing
of  matter systems.

An essential point of the construction \cite{AIMZ} is  the
super-Virasoro constraints at finite $N$ imposed on the model:
\beqn
\label{eq:svir}
  G_{n-1/2}{\bf Z}_{N}[g_{p},\xi_{p+1/2}]  =0~,~~ n=0,1, \cdots \;\;\;.
\eeqn
 The generator $G_{n-1/2}$ is given through the  Neveu-Schwarz supercurrent
 $G(p) = {\displaystyle \sum_{n \in {\cal Z}} }
 \frac{1}{2} G_{n-1/2} p^{-n-1} = \frac{1}{2} \alpha (p) b(p)$ and
 $\alpha (p) = {\displaystyle \sum_{n \in {\cal Z}} } \alpha_{n} p^{-n-1}$
,~~ $b(p) = {\displaystyle \sum_{m \in {\cal Z} +1/2} } b_{m} p^{-m-1/2}$.
 Here, $\alpha_{p} = - \frac{\Lambda}{N} \partial/\partial g_{p},$
$\alpha_{-p}
= - \frac{N}{\Lambda} pg_{p},$ $b_{p+1/2} = - \frac{\Lambda}{N}
 \partial/\partial \xi_{p+1/2},$ and
$b_{-p-1/2} = - \frac{N}{\Lambda } \xi_{p+1/2}$ $p=0,1 \cdots$.
 The super-Vandermonde determinant has been obtained by implementing
 eq.~(\ref{eq:svir}) in the integrand.
  In view of the fact
that the Dyson-Schwinger equation of the ordinary zero-dimensional
 matrix model  is succinctly summmarized as the Virasoro
constraints \cite{FKNDDV}\cite{DAVIDIMITEP} both
 in the double scaling limit \cite{BKDSGM}
 and at finite $N$,
 this construction  albeit being  semi-inductive  can be
 regarded as a logical extension of the bosonic model which
includes supersymmetry.
The model may suggest a new avenue of thoughts toward supersymmetric
triangulations and the combinatorically equivalent matrix integrals
of some kind.

In this letter, we will identify
a super-KP hierarchy attendant with  our model in the double scaling limit.
 The physical significance of this lies in the fact that
 hierarchical equations are able to  relate various correlation functions
 to one another
 and therefore reduce the problem into the one in which only the simplest
 correlator is involved.
Our goal will be accomplished in two steps: first  we take  the double scaling
limit of the super-Virasoro constraints of
 the model at finite $N$. Next, we
 will rederive these double-scaled constraints
from the $\tilde{G}_{-1/2}$ constraint and  two lemmas obtained from
the super-KP hierarchy under consideration
 and its Baker-Akiezer wave functions.  The procedure goes mostly in
parallel to the one discussed in ref. \cite{Goeree}, which owes
some lemmas to ref.~\cite{DJKM}.

Let $Q_{s}$ be a psuedo super-differential operator
\beqn
\label{eq:Q_{s}}
 Q_{s} = D + q_{0} (\{ t_{\ell}\}) +
 \sum_{\ell = 1}^{\infty} q_{\ell}
(\{t_{\ell}\}) D^{-\ell}\;\;,  ~~D \equiv
 \partial/\partial x_{s} + x_{s} \partial/
\partial x \;\;\;, \nonumber
\eeqn
satisfying $Dq_{0}(\{t_{\ell}\}) + 2q_{1}(\{t_{\ell}\}) =0$.
 The coefficients $q_{\ell}$ depend on infinite number of parameters
denoted by $\{ t_{\ell}\}$ or $\{ t \}$.     Here,
$ x=t_{1}, x_{s}= t_{1/2}$.
 It is known that
$Q_{s}$ can then be brought into the  form
 $Q_{s} = K_{s} D K_{s}^{-1}$.
  We will find that  the attendant super-KP
 hierarchy  coincides with
the one proposed by Mulase and Rabin \cite{MulaseRabin},
\beqn
\label{eq:SKP}
\partial K_{s}/\partial t_{\ell} = - \left( Q_{s}^{2\ell}\right)_{-} K_{s}
  \;,\;\;\;
\partial K_{s}/\partial t_{\ell + 1/2} =
  - \left( Q_{s}^{2\ell +1}- K_{s}x_{s} D^{2\ell +2} K_{s}^{-1}
      \right)_{-} K_{s}
  \;\;\;,
\eeqn
confirming the conjecture made in \cite{AIMZ}.
 We denote by $\left( {\cal O}\right)_{+}$ and $\left( {\cal O} \right)_{-}$
 respectively
  the  part containing the non-negative powers  and the one
containing the negative powers of $D$  in  ${\cal O}$.
 We prepare the  Baker-Akiezer wave function
 $ {\bf w}_{\bullet} \left( \lambda, \theta ; \{t_{\ell}\}, \right)$
 (with $\bullet = ({\bf NS})$ or $ ({\bf R})$) which has two different
local expressions:
\beqn
\label{eq:BAW}
  K_{s} e^{ {\displaystyle\sum_{\ell=1}^{\infty}
 t_{\ell} \lambda^{\ell} +\sum_{\ell =0}^{\infty} t_{\ell +1/2} \lambda^{\ell}
 \theta } }~\left( {\bf NS}\right)~  {\rm or}~
  K_{s} e^{  {\displaystyle \sum_{\ell=1}^{\infty}
 t_{\ell} \lambda^{\ell} +\sum_{\ell =0}^{\infty} t_{\ell +1/2} \lambda^{\ell
-1/2} \theta } }~\left( {\bf R} \right)~.
\eeqn
In either case, eq.~(\ref{eq:SKP}) is equivalent to
\beqn
\label{eq:flow}
\partial {\bf w}\left(  \lambda, \theta ;\{t\} \right) /\partial t_{\ell}
 &=& \left( Q_{s}^{2\ell}\right)_{+} {\bf w}\left(
 \lambda, \theta ;  \{ t \} \right)\;,
 \;\;\;   \nonumber  \\
\partial {\bf w}
\left(  \lambda, \theta ; \{t \} \right)/\partial t_{\ell+1/2}
 &=& \left( Q_{s}^{2\ell+1} -K_{s}x_{s}D^{2\ell+2}K_{s}^{-1}
    \right)_{+} {\bf w}\left( \lambda, \theta ;  \{ t \} \right) \;\;\;.
\eeqn

Let us turn to the double scaling limit of
 the superstress tensor $G(p)$. Let $\zeta \equiv p^{2}-1$.
 For simplicity, we will restrict ourselves to
  the planar solution of ref.~\cite{AIMZ} which takes into account
 the even bosonic as well as the even  and odd fermionic couplings
to all orders and
 the odd bosonic couplings  only to the leading order.
   Change of variables we will  make in what follows\footnote{We
 have developed
a systematic method of taking a double scaling limit of the constraints
by making   a change of variables based on the form of the
 scaling operators.  The details will be discussed elsewhere together with
 other points.} is dictated by
 the explicit construction of the scaling operators
 \cite{Kazakov,AIMZ} attendant with this planar solution\footnote{ A duplicity
of the constraints is expected once odd bosonic  couplings are taken to all
 orders. See ref.~\cite{BP}.}.
Let $\alpha(p) = \alpha^{(e-)}(p) +\alpha^{(e+)}(p)
+\alpha^{(o+)}(p) + \alpha^{(o-)}(p) +  N p^{-1}$,
 where the superscripts $e,o,+,-$ denote respectively
 the even, odd,
positive and negative parts of the mode expansion of $\alpha(p)$.
 We have replaced the zero mode part\footnote{This part only renormalizes
a cosmological constant subtractively and will not be discussed hereafter.}
 by the result of its action on the
partition function   ${\bf Z}_{N}$.
Similarly, $ b(p) =  b^{(e-)}(p) + b^{(e+)}(p)+ b^{(o-)}(p)+ b^{(o+)}(p) $.
We will reexpand $ \alpha^{(e-)}(p)$ as ${\displaystyle \sum_{n=0}^{\infty} }
 t_{n}^{B+} \left( -\frac{d}{dp} \sigma_{n}^{B+} \right)$. Here
${\displaystyle
 \sigma_{n-1}^{B+} \equiv -   c_{n} \sum_{k=0}^{n}
\left( \begin{array}{c}
    2n  \\ k \end{array} \right) c_{k}^{-1}  p^{2k}  }$,
{}~with~~
  ${\displaystyle c_{n} \equiv
 \left(-1/4 \right)^{n}  \left( \begin{array}{c}
    2n  \\ n \end{array} \right) }$
is, aside from the normalization, taken from the form of the  scaling
operator \cite{Kazakov}.
 By checking $ [ \alpha^{(e+)}(p), \alpha^{(e-)}(p^{\prime})]$,
one can prove  $\alpha^{(e+)}(p) = {\displaystyle \sum_{n=1}^{\infty}
 \zeta^{-n-1/2} \partial / t_{n-1}^{B+} } $ and
$ \sigma_{n-1}^{B+} = \frac{(-)^{n}}{\pi} {\displaystyle \sum_{\ell=0}
^{n} } B(n-\ell -1/2, 3/2) (- \zeta)^{\ell} \equiv B_{n}(\zeta)$.
The remaining operators  in $\alpha(p), b(p)$
 can be  reexpanded in a similar fashion. We find that
\beqn
\label{eq:reexpand}
\begin{array}{c}
 \alpha^{(e-)} = 2 \left(\frac{\Lambda}{N} \right)^{-1}
\sqrt{1+\zeta} {\displaystyle \sum_{\ell=0}^{\infty} } t_{\ell}^{B+}
 \frac{ dB_{\ell+1}(\zeta)}{d\zeta}\;\;,\;\;
     \alpha^{(e+)} = \left(-\frac{\Lambda}{N} \right)
  {\displaystyle \sum_{\ell=0}^{\infty}} \frac{ \partial}{\partial
 t_{\ell}^{B+} } \zeta^{-\ell -3/2}    \\
\alpha^{(o-)}= 2 \left(\frac{\Lambda}{N} \right)^{-1}
{\displaystyle \sum_{\ell=0}^{\infty} }
 (\ell + \frac{1}{2}) t_{\ell}^{B-} B_{\ell}(\zeta),\;
   \alpha^{(o+)} =   \left(-\frac{\Lambda}{N} \right) \sqrt{1+\zeta}
{\displaystyle \sum_{\ell=0}^{\infty} }
\frac{\partial}{\partial t_{\ell}^{B-} } \zeta^{-\ell -3/2}    \\
  b^{(e-)} =  \left(\frac{\Lambda}{N} \right)^{-1}
 {\displaystyle \sum_{\ell=0}^{\infty} }
  t_{\ell}^{BS+} B_{\ell}(\zeta) \;\;,\;\;
 b^{(e+)} =  \left(-\frac{\Lambda}{N} \right)
 {\displaystyle\sum_{\ell=0}^{\infty}} \frac{\partial}{\partial
t_{\ell}^{BS+} } \zeta^{-\ell-1/2}          \\
    b^{(o-)} =  \left(\frac{\Lambda}{N} \right)^{-1}
 {\displaystyle \sum_{\ell=0}^{\infty} }
 (\ell+\frac{1}{2})t_{\ell}^{BS-} \int^{\zeta}  \frac{d
\zeta^{\prime\prime} }{\sqrt{1 + \zeta^{\prime\prime} } }
B_{\ell}(\zeta^{\prime\prime}),\;
b^{(o+)}= \left(-\frac{\Lambda}{N} \right)  \sqrt{1+\zeta}
{\displaystyle \sum_{\ell=0}^{\infty} }
\frac{ \partial}{\partial t_{\ell}^{BS-}} \times
\end{array}  \nonumber
\eeqn
  $ \zeta^{-\ell-3/2}$ are  consistent
 change of bases appropriate to taking the double scaling
limit.

 We now go on to  derive the double-scaled super-Virasoro constraints.
 Let us imagine evaluating $  \oint \frac{d\zeta}{2\pi i}
 \zeta^{n} G(p)\;,\;\; n =0,1, \cdots$. Inside the $\zeta$ integration,
  the following manipulation is permitted for $B_{\ell} (\zeta)$:
\beqn
\label{eq:Bzeta}
 B_{\ell} (\zeta) = (-)^{\ell} i \oint \frac{d\zeta^{\prime} }{2\pi i}
 \sum_{k=0}^{\infty} \zeta^{\prime (\ell-k-1/2)-1} (1-\zeta^{\prime})^{3/2-1}
(-\zeta)^{k} = - \zeta^{\ell-1/2} (1+\zeta)^{1/2} \;\;\;.  \nonumber
\eeqn
In the first equality, we have used the integral representation of the
Beta function and extended the sum to infinity by analytic continuation.
The original  $\zeta^{\prime}$ contour encloses the origin and one.
In the second equality, we have made the $\zeta^{\prime}$ contour
large ($\mid \zeta^{\prime} \mid > \mid \zeta \mid $ ) to sum the geometric
series and changed integration variables
 to $x \equiv \zeta^{\prime} /\zeta$ and
 $\zeta$ to pick
 a pole at $x=-1$. ( This is allowed as the integrand does not have a cut
 extending from $x=-1$.)
Finally we rescale as
\beqn
\label{eq:rescaling}
\begin{array}{c}
     \zeta =a^{2/m} \zeta_{sc},~~~~
   t_{\ell}^{B\pm} = a^{2(1-\ell/m)} t_{\ell}^{\pm},~~~~
   t_{\ell}^{B S\pm} =  a^{2(1-\ell/m \pm 1/2m)} t_{\ell}^{S\pm} \;\;\;, \\
 \Lambda^{-1} = 1+a^{2}t,~~~~~~ 1/N = \kappa a^{2+1/m} \;\;\;,
\end{array}  \nonumber
\eeqn
 and take $a \rightarrow 0$ limit.
 After some calculation, we find
 ${\displaystyle G(p)
 \stackrel{a \rightarrow 0}{\longrightarrow} \sqrt{2} a^{-3/m}
  \tilde{G} (\zeta_{sc}) }$
  ${\displaystyle 2\tilde{G} (\zeta_{sc}) =
   \sum_{n \in {\cal Z} } \tilde{G}_{n-1/2} \zeta_{sc}^{-n-1}
 = \alpha^{(tw)} \left(\zeta_{sc}\right) \psi_{R}
\left(\zeta_{sc}\right) }$,
\beqn
\label{eq:limitG}
\begin{array}{c}
\alpha^{(tw)} \left(\zeta_{sc}\right)  =
{\displaystyle \sum_{m \in {\cal Z} +1/2}  }
 \alpha_{m} \zeta_{sc}^{-m-1} \;\;, \;\;\; \psi_{R} \left(\zeta_{sc}\right)
 =  {\displaystyle \sum_{m \in {\cal Z} } }
 b_{m} \zeta_{sc}^{-m-1/2} \;\;,   \\
 \alpha_{\ell+1/2} = \left( \frac{\Lambda \kappa}{2} \right)
 \left( \frac{\partial}{\partial t_{\ell}^{+} }
+  \frac{\partial}{\partial t_{\ell}^{-}  } \right)
  \equiv \frac{\partial}{\partial j_{\ell} } \;\;, \;\;
\alpha_{-\ell-1/2} = \left(\ell + \frac{1}{2} \right) \left( \frac{1}
 { \Lambda \kappa} \right)
 \left(
 t_{\ell}^{+} + t_{\ell}^{-} \right)
  \equiv  \left( \ell+ \frac{1}{2} \right) j_{\ell} \;\;,   \\
  b_{\ell} =  \left( \frac{\Lambda \kappa}{\sqrt{2}} \right)
 \left( \frac{\partial}{\partial t_{\ell}^{S+} }
+  \frac{\partial}{\partial t_{\ell-1}^{S-}} \right)
  \equiv  \frac{\partial}{\partial j_{\ell}^{s}} \;\;,  \;\;
b_{-\ell} =  \left( \frac{1}{\sqrt{2}\Lambda \kappa} \right) \left(
 t_{\ell}^{S+} + t_{\ell-1}^{S-} \right)  \equiv j_{\ell}^{s}
 \;\;,  \\
  b_{0} =  \left( \frac{1}{\sqrt{2}\Lambda \kappa}  \right)  t_{0}^{S+}  +
 \left( \frac{\Lambda \kappa}{\sqrt{2}} \right)
 \frac{\partial}{\partial t_{0}^{S+} }
 \;\;, ~~~~ \;\; \ell = 1,2, \cdots \;\;.
\end{array}
 \nonumber
\eeqn
The double-scaled super-Virasoro constraints are expressible in terms of
 the twisted scalar $ \alpha^{(tw)} \left(\zeta_{sc}\right)$
 and the Ramond fermion
$ \psi_{R}\left(\zeta_{sc}\right)$, each of which is made of
 a combination of couplings with
 positive parity and the ones with negative  parity.
 The double-scaled super-Virasoro constraints are stated
as
\beqn
\label{eq:dslsvir}
\begin{array}{c}
 \tilde{G}_{n-1/2} \lim {\bf Z}_{N} =0 \;, \;
 \tilde{G}_{-1/2}= {\displaystyle
   \sum_{\ell=1}^{\infty} }  \left(\ell + \frac{1}{2} \right)
  j_{\ell} \frac{\partial}{\partial j_{\ell}^{s} }
   + {\displaystyle \sum_{\ell =1}^{\infty}  } j_{\ell}^{s}
 \frac{\partial}{\partial j_{\ell-1}}   + \frac{1}{2} b_{0} j_{0}
 \;,   \\
 ~\tilde{G}_{n-1/2}=  {\displaystyle
 \sum_{\ell=0}^{\infty} } \left(\ell + \frac{1}{2} \right)
  j_{\ell} \frac{\partial}{\partial j_{\ell+n}^{s} }
   +  {\displaystyle \sum_{\ell =1}^{\infty} } j_{\ell}^{s}
 \frac{\partial}{\partial j_{\ell+n-1}}   +
  b_{0} \frac{\partial}{\partial j_{n-1}} +
  {\displaystyle \sum_{\ell=0}^{n-2}  }
   \frac{\partial^{2}} {\partial j_{\ell} \partial j_{n-1-\ell}^{s} }
  \;\;.
\end{array}
\eeqn

In the remainder of this letter, we will reobtain eq.~(\ref{eq:dslsvir})
 from the $\tilde{G}_{-1/2}$ constraint and eq.~(\ref{eq:flow}).
We will prove that

\underline{{\it Claim}}:~{\it There exists an operator} ${\cal T}$
{\it  such that the matrix elements (not just the residue) of}
  ${\bf T}_{s} \equiv
  \left[ K_{s} {\cal T} K_{s}^{-1} \right]_{-}$
{\it  vanish once the} $\tilde{G}_{-1/2}$
 {\it  constraint is invoked.}

\noindent
In particular, we will  show that the $ {\bf T}_{s}$ is given by
\beqn
\label{eq:Ttau}
 ~{\bf T}_{s} \delta(x-x^{\prime}) (x_{s}-x_{s}^{\prime})
  = - {\bf {\tau}}^{-1}_{NS} \left[ \{ t \}  \right]
  {\bf {\tau} }^{-1}_{NS}
 \left[ \{ t^{\prime} \}  \right] \times \nonumber \\
 ~res_{\nu, \theta} \left[
  \left(  {\bf V} \left( \nu, \theta; \{  t_{\ell} \} \right)
  S~ res_{\lambda} \left(
\frac{1}{\lambda^{\alpha}} G_{{\rm SKP} }(\lambda) \right) {\bf {\tau} }_{R}
 \left[ \{ t \} \right] \right)
 \left( \tilde{ {\bf V}}
 \left(  \nu, \theta ; \{ t^{\prime} \}  \right)
 {\bf \tau}_{NS}
 \left[ \{ t^{\prime} \}  \right] \right)  \right]
\eeqn
Here, $\alpha = 1/2$ and   $ \{ t \}$ and $ \{ t^{\prime} \}$  differ by
$x \neq x^{\prime}$, $x_{s} \neq x_{s}^{\prime}$ only.  We have introduced
 $ \hat{\bf X}(\lambda, \theta) \equiv
    {\bf X}  (\lambda, \theta ; \{ t \} ) -
     {\bf X}(\lambda^{-1}, \theta ; \{ \partial/ \partial t \} )$, where
\beqn
\label{eq:X}
\begin{array}{cc}
 {\bf X}  (\lambda, \theta ; \{ t \} )=  &
     {\bf X}(\lambda^{-1}, \theta ; \{ \partial/ \partial t \} )=
  \\
 \left\{
\begin{array}{c}
  {\displaystyle\sum_{\ell=1}^{\infty}
 t_{\ell} \lambda^{\ell} +\sum_{\ell =0}^{\infty} t_{\ell +1/2} \lambda^{\ell}
 \theta }   \\
    {\displaystyle \sum_{\ell=1}^{\infty}
 t_{\ell} \lambda^{\ell} +\sum_{\ell =0}^{\infty} t_{\ell +1/2} \lambda^{\ell
-1/2} \theta }
 \end{array} \right.
  \;\;\;  &
   \left\{
\begin{array}{c}
  {\displaystyle\sum_{\ell=1}^{\infty}
  \frac{1}{\ell} \frac{\partial}{\partial t_{\ell}} \lambda^{-\ell}
 -  \sum_{\ell =0}^{\infty} \frac{\partial}{ \partial t_{\ell +1/2} }
 \lambda^{-\ell-1}\theta }\;  ({\bf NS})  \\
  {\displaystyle\sum_{\ell=1}^{\infty}
  \frac{1}{\ell} \frac{\partial}{\partial t_{\ell}} \lambda^{-\ell}
 -  \sum_{\ell = 0}^{\infty}  (1- \frac{1}{2} \delta_{\ell,0})
 \frac{\partial}{ \partial t_{\ell +1/2} }
 \lambda^{-\ell-1/2}\theta   }  ({\bf R})
  \end{array}  \right.
 \end{array}
  \nonumber
\eeqn
 to define
  $ {\bf V} \left( \lambda, \theta ; \{ t \} \right) =
   : e^{ \hat{\bf X}(\lambda, \theta) }:$
  as well as $\tilde{ {\bf V}} \left( \lambda, \theta ; \{ t \} \right)$
   $=: e^{-  \hat{\bf X}(\lambda, \theta)   } : $.
Later, we will  need ${\bf V}  \left( \lambda, \theta,
 \lambda^{\prime}, \theta^{\prime}   ; \{ t \}, \{ t^{\prime} \} \right) =
 : e^{ -\hat{\bf X}(\lambda, \theta) + \hat{\bf X}(\lambda^{\prime},
 \theta^{\prime} ) }: $.
  We denote by $G_{{\rm SKP}}(\lambda) $  a  supercurrent
$\left. - 1/2 {\cal D}\hat{\bf X} (\lambda, \theta)
\partial\hat{\bf X}(\lambda, \theta)
 \right|_{\theta=0}$
 with ${\cal D} \equiv \partial/\partial\theta
 + \theta  \partial/\partial \lambda$.
 The tau function ${\bf {\tau}}_{\bullet} \left[ \{ t \}  \right]$~
(where $\bullet = {\bf (NS)}$ or ${\bf (R)}$) is introduced through
$ {\bf w}_{\bullet}\left( \lambda, \theta;\{t \}\right)$
$= \frac{1}{ \tau_{\bullet} \left[ \{t\} \right] }$
${\bf V}_{\bullet} \left( \lambda, \theta; \{ t\} \right)
 \tau_{\bullet}
 \left[ \{t \}  \right].$
 The  operator $  S \equiv {\displaystyle \lim_{\lambda_{0}
 \rightarrow 0} } S(\lambda_{0})$ is a spin operator \cite{KCFMS} which
creates a cut
 and interpolates between $({\bf NS})$ and $({\bf R})$ sectors
 :$ {\bf {\tau} }_{R}
 \left[ \{t \}  \right] = S  {\bf {\tau} }_{NS}
 \left[ \{t \}  \right]$.\footnote{ The existence
  of such nonlocal operator is strongly supported
 by its appearance in Ising model and the equivalent description
  in terms of the GSO projected
 free Majorana fermion. The operator $S$ establishes a
correspondence  between
$ {\bf {\tau} }_{NS}
 \left[ \{t \}  \right]$ and   ${\bf {\tau} }_{R}
 \left[ \{t \}  \right]$. }

By a straightforward calculation, one  establishes
\beqn
\label{lemma1}
\underline{{\it Lemma 1}}~:~ res_{\nu, \theta}
 \left[  \left(  {\bf P}(\{t\}) e^{x\lambda + x_{s}
\theta} \right) \left(  {\bf R}( \{t^{\prime} \})
 e^{-x^{\prime}\lambda - x_{s}^{\prime}
\theta} \right) \right] ~ =~  \left[  {\bf P}(\{t\})
  {\bf R} ( \{ t^{\prime} \} ) \right]_{-}
 \times \;\;  \nonumber\\
\delta(x-x^{\prime})(x_{s}-x_{s}^{\prime}),{\rm where}~
  {\bf P}( \{t \}) =  \sum_{i= -\infty}^{n}p_{i}
( \{ t \}) D^{i},
{\bf R}( \{ t^{\prime} \}) = \sum_{i^{\prime}
= -\infty}^{m} r_{i^{\prime}}( \{ t^{\prime} \})
 (-D^{\prime})^{i^{\prime}}  \nonumber
\eeqn
   We find  a supersymmetric
 extension of the bilinear identity
 $\int d \lambda d\theta {\bf w}$$( \lambda, \theta; \{t \})$
 ${\bf w}^{*} ( \lambda,$$ \theta;$$ \{ t^{\prime} \})=0$,
 which is a simple consequence
of ${\it Lemma 1}$ and eq.~(\ref{eq:flow}). This can be readily
 seen by expanding the left-hand side of  the equation which we want to prove.
 Let us now prove
\beqn
\label{eq:lemma2}
\underline{{\it Lemma 2}} : \tau^{-1}_{NS} \left[ \{t \}  \right]
  \tau^{-1}_{NS} \left[  \{ t^{\prime} \}  \right]
\left. res_{\nu, \theta} \right[   \left( {\bf V}
 \left( \nu, \theta; \{ t \} \right)
  {\bf V} \left( \lambda, \theta^{\prime \prime}, \mu,
  \theta^{\prime } ;  \{ t \} \right)
  \tau_{NS}
 \left[ \{ t \} \right]  \right) \times\;\;\;
   \nonumber \\
      \left(  \tilde{ {\bf V}}
 \left(  \nu, \theta ; \{ t^{\prime} \}
  \right)    \tau_{NS}
 \left[ \{ t^{\prime} \}  \right] \right)\left] \right.  =
 (\theta^{\prime \prime} -\theta^{\prime})
 {\bf w} ( \mu, \theta^{\prime} ; \{t \})
  {\bf w}^{*} ( \lambda, \theta^{\prime \prime}; \{ t^{\prime} \}) \nonumber \\
 ~  +  \left( \lambda -\mu -\theta^{\prime\prime} \theta^{\prime} \right)
\tau^{-1}_{NS} \left[ \{t \}  \right]
  \tau^{-1}_{NS} \left[  \{ t^{\prime} \}  \right] ~^{\bullet}_{\bullet}
  \left. \right(  {\cal D}{\bf X}( \lambda, \theta^{\prime\prime} ;
\{ t- t^{\prime} \} )   \nonumber \\
 ~-     {\cal D}{\bf X} ( \lambda^{-1}, \theta^{\prime\prime}
 ; \{ \partial/\partial t - \partial/\partial t^{\prime}  \}  )  \left) \right.
   {\bf V}\left( \mu, \theta^{\prime} ;
 \{ t\} \right) \tau_{NS}[ \{ t\}]
 ~\tilde{ {\bf V}}\left( \lambda, \theta^{\prime\prime} ;
 \{ t^{\prime}   \} \right) \tau_{NS} [ \{ t^{\prime}  \}]
   ~^{\bullet}_{\bullet}   \nonumber
\eeqn
  The integrand of the left-hand side  is not normal-ordered.
Putting everything normal ordered, we find that it equals
 ( for $ \mid \nu \mid > \mid \mu \mid, \mid \lambda \mid$)
\beqn
 e^{ {\bf X}(\mu, \theta^{\prime} ; \{ t\} ) - {\bf X}(\lambda,
 \theta^{\prime \prime} ; \{ t\} ) } \left. res_{\nu, \theta} \right[ \left(
e^{- {\bf X}(\nu^{-1}, \theta; \{ \partial/\partial t \} ) +
{\bf X}(\lambda^{-1}, \theta^{\prime \prime}; \{ \partial/ \partial t \} )
-  {\bf X}(\mu^{-1}, \theta^{\prime} ; \{ \partial/\partial t \} ) }
 \tau_{NS}[ \{ t\}] \right) \nonumber \\
 \left( e^{ {\bf X}(\nu^{-1}, \theta; \{ \partial/\partial t^{\prime} \}) }
  \tau_{NS} [ \{ t^{\prime} \}] \right)  e^{  {\bf X}(\nu,
 \theta ; \{ t - t^{\prime} \} ) }  \frac{\nu -\mu -\theta \theta^{\prime} }
{ \nu -\lambda - \theta \theta^{\prime\prime} }  \left] \right.
  \;\;\;. \nonumber
\eeqn
 The bilinear identity ensures that  this expression
has no singularity around the origin. One can, therefore, simply pick
the pole at $\nu = \lambda +\theta \theta^{\prime \prime}$,
  $ \theta = \theta^{\prime \prime}$ to evaluate the integrand. We easily
 find the right hand side. {\it Lemma 2} follows.

 We finally proceed to the computaton of the right hand side
 of eq.~(\ref{eq:Ttau}).
 First, we are concerned with reexpressing the part
$ S~ res_{\lambda} \left[
\frac{1}{\lambda^{\alpha}} G(\lambda)
 \right]{\bf {\tau}}_{R} \left[ \{ t \}  \right] $.
 As the operator product goes as $ G(\lambda) S = \frac{1}{2}
 \lambda^{-3/2} \tilde{S} + o(\lambda^{-1/2}) $,
( $\tilde{S}$ is another spin operator conjugate to $S$ \cite{KCFMS}),
  we find that the  above
expression can be written as
$ \frac{1}{2} res_{\lambda} \left[
\frac{1}{\lambda^{\alpha + 3/2}} {\cal B}(\lambda)
 \right]{\bf {\tau}}_{NS} \left[ \{ t \}  \right] $.
Here ${\cal B}(\lambda)$ is a normal-ordered local  operator
 acting on  the Neveu-Schwarz state  $\tau_{NS}$ \footnote{
 In the bosonic case, the
corresponding operator is simply a stress-energy tensor and explicit
 computation has been given in \cite{Goeree}. This is not the case here.
We will not need it however. }.  We find that any normal-ordered
 local  operator
 consisting of products of $\hat{\bf X}(\lambda, \theta)$
 and its (super)-derivative
 is obtained from
$ {\bf V} \left( \lambda, \theta^{\prime \prime}, \mu,
  \theta^{\prime } ; \{ t \} \right) $ by successive
(super)-differentiations. The right-hand side of eq.~(\ref{eq:Ttau}) can
in principle be evaluated by using  two lemmas estabished, leading to
 a position space expression as given by the left-hand side.
 Let us see this by evaluating the contribution due to $G(\lambda)$.
  Let ${\cal B}(\lambda)$ be $\lambda^{m} G(\lambda) + \cdots$
 with $m$   an integer.
 We find that the right-hand side of eq. (\ref{eq:Ttau}) equals
\beqn
\label{eq:step1}
  - \tau^{-1}_{NS} \left[ \{ t \}  \right]
  \tau^{-1}_{NS} \left[ \{ t^{\prime} \}  \right]
     res_{\lambda} ~  res_{\nu, \theta} [  \left\{
\frac{1}{2\lambda^{\beta}}  \right.
   \left. {\cal D}^{\prime}
\frac{d}{d\mu} {\bf V} \left( \lambda, \theta^{\prime \prime}, \mu,
  \theta^{\prime } ;  \{ t \} \right)
 \right|_{\mu =\lambda, \theta^{\prime}
 = \theta^{\prime\prime} =0}   \nonumber \\
  \left. \left.  \left.
- \frac{\beta }{2\lambda^{\beta +1}}
 {\cal D}^{\prime} {\bf V} \left( \lambda, \theta^{\prime \prime}, \mu,
  \theta^{\prime } ;  \{ t \} \right)
 \right|_{\mu =\lambda, \theta^{\prime}
 = \theta^{\prime\prime} =0}  + \cdots
   \right\}  \tau_{NS}
 \left[ \{ t \}  \right]
 \left( \tilde{ {\bf V}}
 \left(  \nu, \theta ; \{ t^{\prime} \}  \right)
  \tau_{NS}
 \left[ \{ t^{\prime} \}  \right] \right) \right] \;\;\;. \nonumber
\eeqn
 Here, $\beta = 2-m$.
 We can further convert this expression,
 applying formulas obtained by expanding {\it Lemma2} around
 $ (\mu, \theta^{\prime}) = (\lambda, \theta^{\prime \prime}).$
 We find
\beqn
\label{eq:step2}
 \left.  res_{\lambda, \theta^{\prime \prime} } \right[
\frac{\theta^{\prime\prime}}{2\lambda^{\beta }}
 \partial_{ \lambda}{\bf w}
 ( \lambda, \theta^{\prime \prime} ; \{t \})
  {\bf w}^{*} ( \lambda, \theta^{\prime \prime}; \{ t^{\prime} \})
 ~ + \frac{\theta^{\prime\prime}}{2\lambda^{\beta }}
    {\cal D}^{\prime \prime}{\bf w}
 ( \lambda, \theta^{\prime \prime} ; \{t \})
   {\cal D}^{\prime \prime}{\bf w}^{*}
 ( \lambda, \theta^{\prime \prime}; \{ t^{\prime} \})  \nonumber \\
- \frac{ \beta \theta^{\prime\prime}}{2\lambda^{\beta
+1}}{\bf w}
 ( \lambda, \theta^{\prime \prime} ; \{t \})
  {\bf w}^{*} ( \lambda, \theta^{\prime \prime}; \{ t^{\prime} \})
   ~+~  \cdots \left] \right.   \;\;\;. \nonumber
\eeqn
 We use {\it Lemma1} to convert this expression into a position space operator
acting on  $\delta(x-x^{\prime}) (x_{s}-x_{s}^{\prime}) $. In this way,
 we obtain the operator ${\bf T}_{s}$
 $=\left[ K_{s} {\cal T} K_{s}^{-1} \right]_{-}$
such that
\beqn
    {\cal T}  &\equiv& 1/2
 \sum_{\ell=1}^{\infty} \ell t_{\ell} \left( {\partial}/{\partial x}
 \right)^{\ell
-1-\beta} {\partial}/{\partial x_{s}}
+ {1}/{2} \sum_{\ell=0}^{\infty}
 t_{\ell +1/2} \left( {\partial}/{\partial x} \right)^{\ell-\beta}
  \nonumber   \\
  &-& \left( \beta/2 \right)
 \left( {\partial}/{\partial x} \right)^{-1-\beta}
{\partial}/{\partial x_{s}}  + \cdots  \;\;\;.  \nonumber
\eeqn
 We are now convinced that the pseudo-differential operator
 ${\bf T}_{s}$ $=\left[ K_{s} {\cal T} K_{s}^{-1} \right]_{-}$
 in
 Eq.~(\ref{eq:Ttau})  exists.

 The two reduction of the hierarchy  is stated  as
\beqn
\label{eq:reduction}
  (\partial/\partial t_{2\ell})   \tau
 \left[ \{ t \}  \right]  =0 \;\;,\;\;\;
  (\partial/\partial t_{2\ell- 1/2})   \tau
 \left[ \{ t \}  \right] =0 \;\;,\;\; \ell = 1,2, \cdots
\eeqn
We observe that
$ \frac{1}{\sqrt{2}} res_{\lambda} \left(
\frac{1}{\lambda^{\alpha}} G_{{\rm SKP}}(\lambda) \right) {\bf {\tau} }_{R}
 \left[ \{ t \} \right] $ coincides with $\tilde{G}_{-1/2}
 {\displaystyle \lim_{ a \rightarrow 0} }{\cal Z}_{N}$
if the  identification
\beqn
\label{eq:ident}
  \sqrt{2} t_{2\ell+1} =j_{\ell} \; (\ell =0,1, \cdots)\;\;,
\;\; t_{2\ell+1/2} =j_{\ell}^{s} \; (\ell = 1,2, \cdots)\;\;, \;\;
\left(\sqrt{2}\Lambda \kappa \right) t_{1/2} =  t_{0}^{S+}
  \nonumber
\eeqn
is made
 and if the partition function is identified as
 the $\tau$ function. Eq.~(\ref{eq:Ttau}), therefore, implies
 the vanishing of the matrix elements of ${\bf T}_{s}$
 under the $\tilde{G}_{-1/2}$ constraint. The {\it Claim} has now been
proven.

Let $\hat{\lambda} = K_{s} \frac{\partial}{\partial x} K^{-1}_{s}$.
We find, under the constraint,
\beqn
\label{eq:rest}
 \left( \hat{\lambda}^{2\ell} K_{s} {\cal T} K_{s}^{-1} \right)_{(-)}
     = \hat{\lambda}^{2\ell} {\bf T}_{s} =0 \;\;\;
\eeqn
 as $\hat{\lambda}^{2}$ is a differential ( as opposed to a
 pseudo-differential)
 operator under eq.~(\ref{eq:reduction}).
On the other hand,
$ \left( \hat{\lambda}^{2\ell} K_{s} {\cal T} K_{s}^{-1} \right)_{(-)}
\delta(x-x^{\prime}) (x_{s}-x_{s}^{\prime})$ is  equal to the right-hand
side of eq.~(\ref{eq:Ttau}) with $\alpha (=1/2)$ replaced by $\alpha-2\ell$.
We conclude that eq.~(\ref{eq:rest}) implies
\beqn
\label{eq:higherG}
   res_{\lambda} \left(
\lambda^{-\alpha+2\ell} G_{\rm SKP}(\lambda) \right) {\bf {\tau} }_{R}
 \left[ \{ t \} \right]  =0 \;\;\;.  \nonumber
\eeqn
But this, together with eq.~(\ref{eq:reduction}),
 is nothing but the ${\tilde G}_{\ell-1/2}$ constraint
 in eq.~(\ref{eq:dslsvir}), which we wanted to rederive.

It is straightforward to extend this proof to the $p$-th reduction of
the super KP hierarchy discussed here
 and will be reported elsewhere.

The author thanks  Luis Alvarez-Gaum\'{e} and Juan Ma\~{n}es for continuing
discussions on this subject. They have independently obtained the double
scaling limit of the super-Virasoro constraints.  The author has been informed
that they carried out an explicit computation up to genus two  together
with K. Becker and M. Becker \cite{AM}.

\end{document}